\documentclass[fleqn,12pt,twoside]{article}
\usepackage{espcrc1}

\usepackage{graphicx}

\title{Neutrino beams as a probe of the nuclear isospin and spin-isospin
excitations}

\author{R.~Lazauskas\address[IPN]{Institut de Physique Nucl\'eaire,
F-91406 Orsay cedex, France}\thanks{E-mail: lazauskas@lpsc.in2p3.fr} ,C. Volpe{\addressmark[IPN]}\thanks{E-mail: volpe@ipno.in2p3.fr}}

\begin{document}

\maketitle

\begin{abstract}
We explore the possibility of performing nuclear structure studies using low energy
neutrino beams.
In particular, low energy beta-beams and conventional sources (muon
decay-at-rest) are considered. We present results on the total
charged-current as well as flux-averaged cross sections associated to
electron (anti)-neutrino scattering on oxygen, iron, molybdenum and lead, as
typical examples. It is shown that by using neutrinos from low energy
beta-beams, information on forbidden states, in particular the spin-dipole,
could be extracted.
\end{abstract}

\section{Introduction}

\noindent Nuclear isospin and spin-isospin excitations play a crucial role
in weak processes.
A precise description of these transitions is necessary to
progress on open issues in high energy physics and in
astrophysics. Searches for processes beyond the Standard Model,
such as neutrinoless double beta-decay, require a detailed
knowledge of the isospin and spin-isospin nuclear response. In
particular, beta-decay \cite{betadecay}, muon capture \cite%
{Kortelainen:2002bz,Kortelainen:2004gr} the $2\nu$ double-beta decay
\cite{Rodin:2003eb}
as well as charge-exchange studies
\cite{Akimune:1997qu,Rakers:2004uk,Rakers:2005ft} are essential to
constrain neutrinoless double-beta decay predictions; while neutrino-nucleus
has been pointed out recently as a new constraint \cite{Volpe:2005iy}. Beta decay
is one of the key ingredients of the nucleosynthesis of heavy elements
(see e.g. \cite{Engel:1999tr,Borzov:2000ve}). The role of neutrino-nucleus
interactions is emphasized in  \cite{Heger:2003mm}  for
neutrino-nucleosynthesis and for the r-process \cite{Meyer:1998sn,Surman:1998eg,Langanke:2002ab,Balantekin:2003ip}.
Such processes are also an essential ingredient to determine the
response of neutrino detectors which can  serve as core-collapse
Supernova observatories.

The Isobaric Analogue States (IAS) as well as the Gamow-Teller
(GT) giant resonances represent the most studied cases.
Information on the associated nuclear transitions is essentially
obtained through beta-decay or charge-exchange reactions. A
satisfactory description of these excitations is now achieved
using microscopic approaches: the shell model or the
random-phase-approximation (RPA) and its variants. The
effective interactions associated with these models have been
extensively discussed and some of their parameters have been
determined using the allowed transitions as a constraint (see e.g. \cite{Bender:2001up}).
However, one is still not able to explain why the measured GT
transition amplitudes are always larger than the calculated ones.
This is usually corrected by using an effective axial-vector
coupling constant (the "quenching problem" \cite{Osterfeld:1991ii,Caurier:2004gf}).

Little is still known for the states of higher multipolarity,
such as the spin-dipole or the higher multipoles. Their
importance has been underlined in astrophysics (see e.g. \cite{Surman:1998eg}%
) and is well established for neutrinoless double-beta decay (see e.g. \cite%
{multipole,Civitarese:2005jf}).  Using sum-rules, one can relate
the spin-dipole strength distribution to the neutron skin thickness defined as the difference between the root mean square radii of the proton and neutron  distributions. This is
an useful information to establish the nuclear matter equation of state of neutron stars \cite{Yako:2006gz}.
Some information on these states has been extracted thanks to various weak and hadronic probes.
A systematic study of muon capture is performed in \cite{Zinner:2006jv} and it is shown that
no quenching of the forbidden states is required to account for the total capture
rates. On the other hand, experiments with (p,n) and ($^{3}$He,t) have been realized
\cite{Krasz:PRL99}. Calculations based on RPA overestimate the experimental
results in the Gamow-Teller region, while they underestimate them
in the 30-60 MeV energy range \cite{Yako:2006gz}.
Therefore if quenching of the forbidden states is necessary is an open issue.
This question could be settled if more experimental information was available. 
Furthermore the latter could be exploited to improve the parametrization of the phenomenological interactions used in microscopic approaches.

Neutrino-nucleus interactions can probe different spin-isospin and
isospin states, depending on the neutrino energy. While neutrinos
in the tens of MeV, such as those produced by the sun, essentially
explore the IAS and GT excited states, those having several tens
to hundreds MeV, e.g. from core-collapse supernovea or from
accelerators, also probe the spin-dipole and higher multipoles.
These measurements are in most cases inclusive. However, different
pieces of information can be obtained by using either sources
where the energy of the neutrinos can be tuned or complementary
probes. So far experiments are scarce, and include: a measurement
on deuteron using reactor neutrinos \cite{Willis:1980pj}, one on
iron \cite{Zeitnitz:1998qg} and several on carbon with
conventional sources (muon decay-at-rest or decay-in-flight)
\cite{expc12}. Therefore for the numerous applications, one relies
on the theoretical predictions available in the literature and
based on different approaches. The calculations generally agree
for neutrinos in the ten of MeV where the response is dominated by
the IAS and the GT transitions. However, significant discrepancies
appear at higher energies (see e.g. \cite{c12,Jachowicz:2002rr}
for carbon and
\cite{Fuller:1998kb,Kolbe:2000np,Volpe:2001gy,SajjadAthar:2005ke}
for lead). The origin of these disagreements is not yet clear and
can be due to a different choice in the nuclear structure
description (e.g. the continuum treatment, the choice of the
effective forces, or the correlations included).

Systematic neutrino-nucleus interaction measurements could be an ideal tool
to explore the weak nuclear response. At present, new experiments on various
nuclei are being proposed with a new facility using muon decay-at-rest \cite%
{Avignone:2003ep}. Another possibility could be offered by beta-beams. This
is a new method to produce pure and well known electron neutrino beams,
exploiting the beta-decay of boosted radioactive ions \cite{Zucchelli:2002sa}%
. The idea of establishing a low energy beta-beam facility has been first
proposed in \cite{Volpe:2003fi}. Its potential has been discussed in \cite%
{Volpe:2003fi,Serreau:2004kx,McLaughlin:2004va} for nuclear structure
studies, in \cite{Volpe:2003fi,Jachowicz:2006xx} for core-collapse supernova physics, and
in \cite{McLaughlin:2003yg,Volpe:2005iy,Balantekin:2005md,Balantekin:2006ga,Bueno:2006yq,Barranco:2007tz}
for the study of fundamental interactions. (For a review on the low, medium and high
energy scenarios as well as their physics potential see \cite{Volpe:2006in}).
An analysis of how neutrino spectral shapes change at low-energy
beta-beams depending on the detector geometry and different locations within
the same detector is made in \cite{Amanik:2007zy}, while the possibility of exploiting
low-energy neutrinos at off-axis from a standard beta-beam is proposed in \cite{Lazauskas:2007va}.

In this paper, we consider low energy beta-beams as a tool to
study the isospin and spin-isospin nuclear response. This feature
has been first underlined in \cite{Volpe:2003fi} and explored for
the neutrino-lead reaction in \cite{McLaughlin:2004va}. Here we
perform microscopic calculations of charged-current
neutrino-nucleus interactions. The nuclear amplitudes are obtained
using the mean field approach including pairing correlations and
the random-phase-approximation among quasi-particles. We present
the total neutrino-nucleus cross sections as well as the
contribution of the various multipoles and discuss how their
importance evolve, as a function of neutrino energy. We give the
flux-averaged cross sections associated to low energy beta-beams
and to conventional sources (muon decay-at-rest), and compare the
latter to previous calculations, when available. Results are given
for oxygen, iron, molybdenum and lead taken as sample nuclei.
These nuclei present particular interest for neutrino detectors
and supernova observatories, currently used or under study. In
particular, oxygen is used in water Cherenkov detectors, like
Super-Kamiokande, the next-generation UNO
\cite{uno}, MEMPHYS \cite{deBellefon:2006vq} and Hyper-Kamiokande \cite{itow}%
. An iron-based detector was considered in \cite{Avignone:2003ep}; while
molybdenum- \cite{Ejiri:1999rk} and lead \cite{elliott} ones are being
studied. Note that the contribution of the different multipoles
to the neutral current neutrino-lead cross section
was studied theoretically in Ref.\cite{Jachowicz:2002hz}.

The paper is organized as follows. The theoretical aspects are described in
Section II. Results on the isospin and spin-isospin states contributing to
the cross sections as well as on the total and the flux-averaged cross
sections are presented in Section III. Conclusions are drawn in Section IV.

\section{Calculations}

\noindent The total cross section for the charged current
neutrino-nucleus scattering process $\nu _{l} (\bar{\nu
}_{l})+^{A}X \rightarrow l+^{A}Y$ is given by \cite{Walecka}
\begin{equation}  \label{e:1}
\sigma (E_{\nu}) =-i\int d^{3}k_{l}\delta (E_{l}+E_{f}-E_{\nu
}-E_{i})|\langle l(\vec{k}_{l});f|\mathit{H_{eff}}|\nu
_{l}(\vec{k}_{\nu});i\rangle |^{2},
\end{equation}
where $E_{f}$ ($E_{i}$) is the energy of the final (initial)
nuclear state, $E_{\nu }$ ($\vec{k}_{\nu }$) is the incident
neutrino energy (momentum) and $E_{l}$ ($\vec{k}_{l}$)
is the outgoing lepton energy (momentum). The effective
single-particle hamiltonian $\mathit{H_{eff}}$ is derived by
carrying out the Foldy-Wouthuysen (FW) transformation for the
nucleon weak current operator and retaining terms up to
$O(|{\mathbf{q}}|/M$) (M is the nucleon mass and ${\mathbf{q}}$ is
the momentum transfer). After performing a multipole expansion of
the weak current, the differential cross section for the
transition between nuclear states with angular momenta  $J_{i}$
and $J_{f}$ is given by:
\begin{equation} \label{e:2}
\frac{d\sigma(E_{\nu}) }{d\Omega }={%
\frac{G^{2}cos^{2}\theta _{C}E_{l}k_{l}}{{\pi }}}\frac{1}{2J_{i}+1}\left(
\sum_{J=0}^{\infty }\sigma _{CL}^{J}+\sum_{J=1}^{\infty }\sigma
_{T}^{J}\right)
\end{equation}%
where $G\,cos\,\theta _{C}$ is the weak coupling constant. The
$\sigma _{CL}^{J}$ and $\sigma _{T}^{J}$ represent the
longitudinal and traverse response of the system to the external
field. According to~\cite{Walecka} these can be expressed as
follows:

\begin{eqnarray}\label{e:3}
\sigma _{CL}^{J} &=&(1+\widehat{\upsilon }\mathbf{\cdot \beta })\left\vert
\left\langle J_{f}\left\Vert \widehat{\cal{M}}_{J}\right\Vert
J_{i}\right\rangle \right\vert ^{2}+\left[ 1-\widehat{\upsilon }\mathbf{%
\cdot \beta }+2\left( \widehat{\upsilon }\mathbf{\cdot }\widehat{q}\right)
\left( \widehat{q}\mathbf{\cdot \beta }\right) \right] \left\vert
\left\langle J_{f}\left\Vert \widehat{\cal{L}}_{J}\right\Vert
J_{i}\right\rangle \right\vert ^{2} \\ \nonumber
&&- \left[ \widehat{q}\mathbf{\cdot }\left( \widehat{\upsilon }\mathbf{+\beta
}\right) \right] 2Re\left\langle J_{f}\left\Vert \widehat{\cal{L}%
}_{J}\right\Vert J_{i}\right\rangle \left\langle J_{f}\left\Vert \widehat{%
\cal{M}}_{J}\right\Vert J_{i}\right\rangle ^{\ast }
\end{eqnarray}

\begin{eqnarray}\label{e:4}
 \sigma _{T}^{J} &=&\left[
1-\left( \widehat{\upsilon }\mathbf{\cdot }\widehat{q}\right)
\left( \widehat{q}\mathbf{\cdot \beta }\right) \right] \left(
\left\vert \left\langle J_{f}\left\Vert
\widehat{\cal{J}}_{J}^{mag}\right\Vert J_{i}\right\rangle
\right\vert ^{2}+\left\vert \left\langle J_{f}\left\Vert
\widehat{\cal{J}}_{J}^{el}\right\Vert J_{i}\right\rangle
\right\vert
^{2}\right)  \\ \nonumber
&&\pm \left[ \widehat{q}\mathbf{\cdot }\left( \widehat{\upsilon }\mathbf{%
-\beta }\right) \right] 2Re\left\langle J_{f}\left\Vert \widehat{%
\cal{J}}_{J}^{mag}\right\Vert J_{i}\right\rangle \left\langle
J_{f}\left\Vert \widehat{\cal{J}}_{J}^{el}\right\Vert
J_{i}\right\rangle ^{\ast }
\end{eqnarray}

\noindent where $\widehat{\upsilon }=\vec{k}_{\nu }/E_{\nu }$ is a
unit vector defining the incoming (anti)neutrino flux
$;\mathbf{\beta =}\vec{k}_{l}/E_{l}
$.  $\widehat{\cal{M}}_{J},$ $\widehat{\cal{L}}_{J},$ $\widehat{%
\cal{J}}_{J}^{el}$ and $\widehat{\cal{J}}_{J}^{mag}$ are
the charge, longitudinal, traverse electric and traverse magnetic
multipole operators, respectively. The plus or minus signs
correspond to neutrino or anti-neutrino scattering respectively.
The corresponding matrix elements defined for single-particle
transitions can be found in~\cite{Walecka}.

A correction to~(\ref{e:2}) must be introduced to account for the
distortion of the outgoing lepton wave function due to the Coulomb
field of the daughter nucleus. While the Fermi function
$F(Z_{f},E_{l})$ accounts well for it at low
neutrino energies, the modified ``Effective Momentum
Approximation'' (EMA) offers a good procedure at high
neutrino energies~\cite{Engel:1997fy}. Here we follow the prescription
already used in previous calculations, where the Fermi
function is taken until the corresponding corrections give a
larger cross section than those obtained with the EMA
approximation.

Flux-averaged cross sections are the measured observables. These
are obtained by folding the energy dependent cross
section~(\ref{e:2}) with the neutrino flux $\phi(E_{\nu})$ that
depends on the specific neutrino source and the target geometry

\begin{equation}\label{e:7}
\langle \sigma \rangle_{\phi} = \int dE_{\nu} \sigma(E_{\nu}) \phi(E_{\nu}).
\end{equation}

The transition matrix elements entering in
Eqs.~(\ref{e:3}-\ref{e:7}) are calculated within microscopic
approaches. The starting point is a Hartree-Fock (HF) calculation
for the ground-state of the nucleus, performed in coordinate space
by using the Skyrme-type effective interactions. The HF procedure
determines the mean-field and single-particle (s.p.) occupied
levels. The unoccupied levels are obtained by diagonalizing the HF
mean-field using a harmonic oscillator basis. Therefore, the
continuum part of the s.p. spectrum is discretized and discrete
particle-hole (ph) configurations coupled to a given $J^\pi$ are
used as a basis, in order to cast the RPA equations in the matrix
form.
This RPA calculation is self-consistent since the residual
interaction among ph states is derived from the same Skyrme force
as the one used to produce the mean field.

When necessary, to go beyond the closed-shell approximation for the ground
state, pairing correlations are taken into account in the HF+BCS
approximation. Constant pairing gaps $\Delta_p$ and $\Delta_n$ for protons
and neutrons are introduced. On top
of the HF+BCS calculation, the QRPA matrix equations can be written with a
procedure which parallels what was described above, with the
two-quasiparticle (2qp) configurations replacing the ph ones. We do not
present here the details of the QRPA formalism, which can be found in the
literature. We simply note that the
particle-particle matrix elements are here renormalized by means of a
parameter $g_{\mathrm{pp}}$ that has been chosen to be smaller than 1
(typically 0.7) to avoid the well-known ground-state instabilities.

For each multipolarity, every eigenstate of the RPA or QRPA equations is
characterized by its $X^f$ and $Y^f$ amplitudes and the transition matrix
element for a generic operator $\hat{O}(k)$ is written as
\begin{equation}
\langle {J_f \vert \sum_k \hat O(k) \vert \mathit{J_i}} \rangle =
\sum_{\alpha,\beta} \langle \alpha \vert \sum_k \hat O(k) \vert \beta
\rangle (X^f_{\alpha\beta}u_\alpha v_\beta + Y^f_{\alpha\beta}v_\alpha
u_\beta),
\end{equation}
where $\alpha$ and $\beta$ label a given ph or 2qp states, $u$ and $v$ are
the BCS occupation amplitudes (which reduce to 1 and 0 in the HF-RPA case)
and $\langle \alpha \vert \sum_k \hat O(k) \vert \beta \rangle$ are
single-particle matrix elements.

\section{Results and discussion}

\begin{figure}[tbp]
\begin{center}
\includegraphics[scale=0.9]{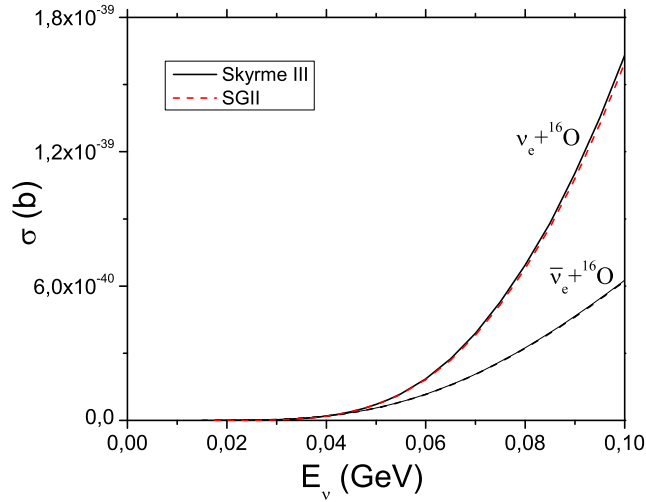}
\end{center}
\caption{Total neutrino-$^{16}$O cross section as a function of neutrino
energy. Results obtained with the SGII and Skyrme III forces are shown.}
\label{fig:totoxygen}
\end{figure}

\begin{figure}[tbp]
\begin{center}
\includegraphics[scale=0.9]{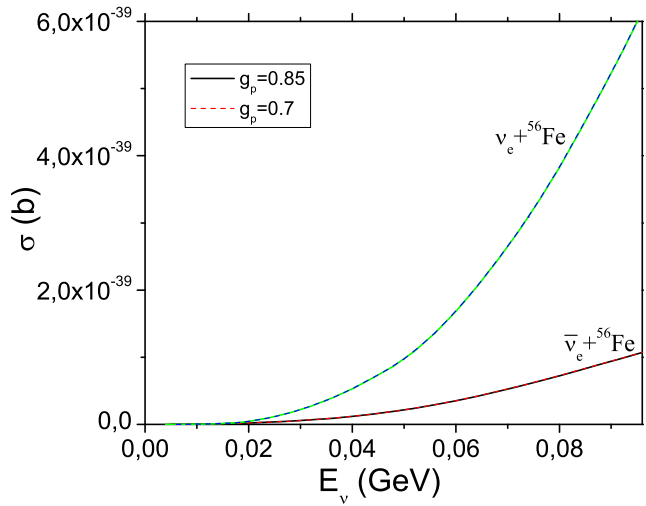}
\end{center}
\caption{Total (anti)neutrino-$^{16}$Fe cross section as a function of
neutrino energy. Results obtained for two values of the $g_{pp}$ parameter
renormalizing the particle-particle interaction in the QRPA calculations (see text) are given.}
\label{fig:totfer}
\end{figure}

\noindent The calculations we show are performed using the Skyrme force. We have
checked that the use of the SGII force does not change the results significantly (Figure \ref{fig:totoxygen}).
The value of $g_{pp}=$0.7 is used, but the results are stable against
changes of this parameter (Figure \ref{fig:totfer}). All states up to $J^{\pi}=5^{\pm}$ are included.
We present results for four nuclei as typical examples, namely $^{16}$O, $%
^{56}$Fe, $^{100}$Mo and $^{208}$Pb. The total (anti-)neutrino cross sections  are given in Table \ref{table:totalcross} for the four nuclei considered.

We have compared our results to previous calculations, when available,
either on the total cross sections as a function of neutrino energy, or on
the flux-averaged cross sections associated to electron neutrinos from
decay-at-rest muons. Our results are in reasonable agreement with those of Refs.
\cite{Jachowicz:2002rr,SajjadAthar:2005ke,Kolbe:2002gk} for oxygen, of Ref.~\cite%
{Kolbe:2000np,SajjadAthar:2005ke,Kolbe:1999vc,Mintz:2002rt} for
iron and of Ref.\cite{Kolbe:2000np,SajjadAthar:2005ke,sagawa} for
lead. As far as the comparison with measurements is concerned, the
only available is the neutrino-iron cross section. The muon
decay-at-rest flux averaged cross section is $256 \pm 108 \pm 43$
($10^{-42}$cm$^{2}$). Our result, 352 $10^{-42}$cm$^{2}$,  is in
agreement with the experimental value.

\begin{figure}[tbp]
\begin{center}
\begin{minipage}{8.2 cm}
\includegraphics[width=9.cm]{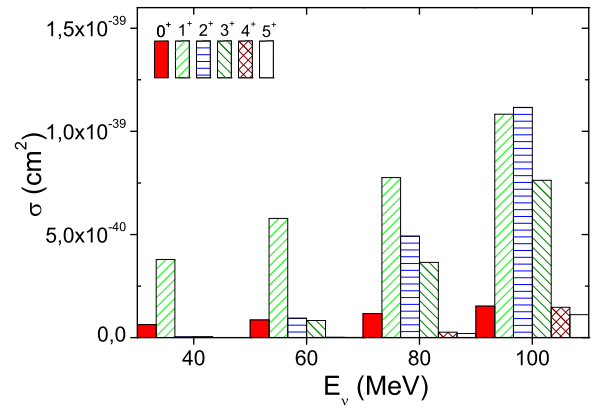}
\end{minipage}
\begin{minipage}{8.2 cm}
\includegraphics[width=9.cm]{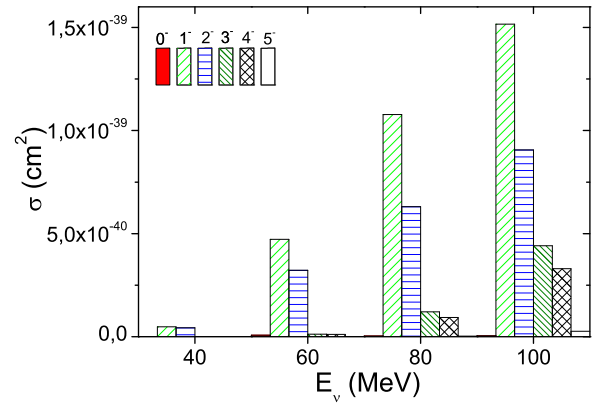}
\end{minipage}
\end{center}
\par
\caption{Contribution of the positive (left) as well as negative (right) parity states to
the total neutrino-iron cross section, as a function of neutrino energy. }
\label{fig:muliron}
\end{figure}

The contribution of the different multipoles as a function of the
neutrino energy are shown in Figure~\ref{fig:muliron} for iron, as
an example, both for positive and negative parity states. Below 40
MeV the GT $J^{\pi}=1^{+}$ transitions dominate the cross section.
At 60 MeV the $J^{\pi}=1^{-},2^{-}$ become important as well.
Beyond 80 MeV all states contribute.

\begin{figure}[tbp]
\begin{center}
\includegraphics[scale=0.7]{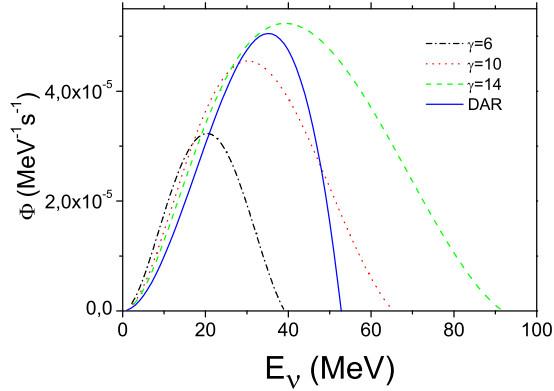}
\end{center}
\caption{Anti-neutrino fluxes from the decay of $^{6}$He ions boosted at $%
\protect\gamma=6$ (dot-dashed line),$\protect\gamma=10$ (dotted line) and $%
\protect\gamma=14$ (dashed line). The full line presents the Michel spectrum for
neutrinos from muon decay-at-rest. }
\label{fig:lownu}
\end{figure}

The details on the neutrino flux associated to low-energy
beta-beams are given in \cite{Serreau:2004kx}
(Fig.~\ref{fig:lownu}).
Following a preliminary feasibility study, we assume, as in~\cite%
{Balantekin:2005md,Balantekin:2006ga}, that the boosted ions are
stored in a storage ring which has 1885 m total length and 678 m
straight sections, and
that the detectors are located at 10 m from the storage ring~\cite%
{Chance:0000}. We consider fully efficient cylindrical detectors
as in~\cite{Serreau:2004kx}, namely $R=1.5$ m (radius) and $h=4.5$
m  (depth) for metals and $R=4.5$ m, $h=15$ m  water Cherenkov
detectors for oxygen. The neutrino flux corresponding to
conventional neutrino sources, i.e. muon decay-at-rest (DAR) is
given by the well known Michel spectrum. As seen in
Figure~\ref{fig:lownu}, the latter has quite similar shape to a
low energy beta-beam with $\gamma=10$.
Note that in principle since the cross sections approximately grow as the neutrino 
energy square, the flux-averaged cross sections can show differences
due to the high energy part of the neutrino flux, as discussed in detail for the case
of core-collapse supernova neutrino spectra \cite{Jachowicz:2003iz}. 

Table~\ref{tab:fluxavernu} presents the contribution (in
percentage) of the different states to the flux-averaged cross
sections Eq.(\ref{e:1}-\ref{e:7}). One can see that in all cases
the results for $\gamma=10$ are similar to the DAR case. The
neutrino-oxygen cross section is dominated by the $1^{-},2^{-}$.
Only with increasing $\gamma$, the relative contribution of the
$2^{-}$ decreases in favor of the positive parity multipoles. 
This behaviour is due to the oxygen structure:  positive parity transitions
require at least 2$\hbar \omega$ excitations. 
The situation
is different for the neutrino-iron, neutrino-molybdenum and
neutrino-lead cross sections. For these nuclei the $\gamma=6$ case
is dominated by the $0^{+},1^{+}$ transitions, with a relative
ratio of about 1/3. When the ion boost increases, their
contribution diminishes basically in favor of $1^{-},2^{-}$. For
$\gamma=14$ the $2^{+}$ and $3^{+}$ contribution become as
important as the Isobaric Analogue state.
Table~\ref{table:fluxaveranu} presents the results for the
anti-neutrino scattering where the anti-neutrino fluxes are
produced by the decay of boosted $^{6}$He ions.
While the $1^{+}$
dominates the flux-averaged cross section at $\gamma=6$ for the
iron nucleus, about 50$\%$ of its value is given by the
$1^{-},2^{-}$ transitions at ion boost 14. On the contrary, these
states dominate at low gamma for the lead nucleus, while almost
all multipoles contribute at high gamma. 
This behaviour can be well understood in terms of the underlying
shell structure of the lead nucleus, 
since the lowest pn-excitations between the filled and
unfilled shells are of negative parity. 

\begin{table}[tbp]
\begin{center}
\begin{tabular}{|lllllll|}\hline
$E_\nu $(MeV) & $\nu$-$^{208}$Pb & $\overline{\nu}$-$^{208}$Pb &
$\nu$-$^{100}$Mo& $\nu$-$^{56}$Fe       &  $\overline{\nu}$-$^{56}$Fe &
$\nu$-$^{16}$O     \\
\hline
7.5       &  2.47(-4)&   1.37(-6)&  2.30(+0) &  8.77(-1) &  8.41(-1) & -   \\
10.0      &  8.49(0) &  8.36(-3) &  6.84(+0) &  3.63(+0) &  2.95(+0) & - \\
12.5      &  4.68(+1)&  7.28(-2) &  1.97(+1) &  8.94(+0) &  6.09(+0) & - \\
15.0      &  1.75(+2)&  2.44(-1) &  4.60(+1) &  1.73(+1) &  1.03(+1) & - \\
20.0      &  8.53(+2)&  1.11(0)  &  2.09(+2) &  5.26(+1) &  2.17(+1) &
7.11(-2)  \\
25.0      &  2.86(+3)&  3.05(0)  &  5.01(+2) &  1.25(+2) &  3.74(+1) &
5.27(-1)   \\
30.0      &  4.90(+3)&  1.53(0)  &  9.04(+2) &  2.33(+2) &  5.74(+1) &
2.48(0)   \\
40.0      &  7.13(+3)&  5.65(0)  &  1.96(+3) &  5.44(+2) &  1.20(+2) &
1.91(+1)  \\
50.0      &  1.13(+4)&  3.48(+1) &  3.02(+3) &  9.83(+2) &  2.09(+2) &
7.13(+1)\\
60.0      &  1.63(+4)&  8.29(+1) &  4.67(+3) &  1.67(+3) &  3.53(+2) &
1.85(+2)\\
70.0      &  2.20(+4)&  1.46(+2) &  6.80(+3) &  2.59(+3) &  5.26(+2) &
3.89(+2) \\
80.0      &  2.83(+4)&  2.16(+2) &  9.36(+3) &  3.73(+3) &  7.27(+2) &
6.94(+2)  \\
90.0      &  3.50(+4)&  2.91(+2) &  1.23(+4) &  5.07(+3) &  9.40(+2) &
1.10(+3)  \\
100.0     &  4.16(+4)&  3.67(+2) &  1.55(+4) &  6.60(+3) &
1.14(+3)& 1.63(+3)
\\ \hline
\end{tabular}
\end{center}
\par
\vskip 0.5cm
\caption{Total cross sections for the indicated
neutrino nucleus charged-current reactions as a function of incoming neutrino
energy. The cross sections are given in units of 10$^{-42}$~cm$^2$,
exponents are given in parentheses.} \label{table:totalcross}
\end{table}

\begin{table}[tbp]
\begin{center}
\begin{tabular}{|c|c|cccc|cccc|}
\hline
\multicolumn{1}{|c|}{$~^{A}$X~} & \multicolumn{1}{c|}{$~~$} &
\multicolumn{1}{c}{$~0^{+}~$} & \multicolumn{1}{c}{$~1^{+}~$} &
\multicolumn{1}{c}{$~2^{+}~$} & \multicolumn{1}{c|}{$~3^{+}~$} &
\multicolumn{1}{c}{$~0^{-}~$} & \multicolumn{1}{c}{$~1^{-}~$} &
\multicolumn{1}{c}{$~2^{-}~$} & \multicolumn{1}{c|}{$~3^{-}~$} \\ \hline
$^{16}$O & $\gamma=6$ & 0.17 & 0.71 & 0.16 & 0.15 & 0.97 & 34.1 & 63.8 & 0.03
\\
& 10 & 0.28 & 1.90 & 1.21 & 1.25 & 0.65 & 43.2 & 51.3 & 0.18 \\
& 14 & 0.43 & 4.04 & 3.36 & 3.33 & 0.34 & 42.4 & 44.8 & 0.75 \\
& $DAR$ & 0.23 & 1.38 & 0.70 & 0.74 & 0.79 & 42.3 & 53.7 & 0.10 \\
$^{56}$Fe & 6 & 22.6 & 73.1 & 0.16 & 0.19 & 0.03 & 1.6 & 2.32 & 0.0 \\
& 10 & 11.2 & 59.5 & 2.04 & 1.91 & 0.30 & 13.8 & 10.9 & 0.18 \\
& 14 & 6.44 & 39.0 & 6.65 & 5.43 & 0.26 & 23.7 & 15.8 & 1.25 \\
& $DAR$ & 13.0 & 66.7 & 1.09 & 1.12 & 0.24 & 9.45 & 8.27 & 0.06 \\
$^{100}$Mo & 6 & 22.6 & 69.7 & 0.40 & 0.39 & 0.07 & 2.61 & 4.21 & 0.01 \\
& 10 & 14.6 & 47.3 & 3.24 & 3.16 & 0.28 & 15.6 & 14.5 & 0.66 \\
& 14 & 9.00 & 31.3 & 7.97 & 6.68 & 0.18 & 21.5 & 17.3 & 2.70 \\
& $DAR$ & 16.7 & 53.3 & 2.12 & 2.08 & 0.30 & 12.2 & 12.6 & 0.29 \\
$^{208}$Pb & 6 & 21.0 & 62.2 & 1.77 & 1.50 & 0.08 & 2.88 & 10.1 & 0.20 \\
& 10 & 11.8 & 39.3 & 6.63 & 5.91 & 0.24 & 15.1 & 16.8 & 1.82 \\
& 14 & 7.77 & 26.6 & 11.2 & 8.81 & 0.12 & 17.0 & 15.7 & 5.06 \\
& $DAR$ & 12.9 & 43.3 & 4.95 & 4.76 & 0.32 & 14.1 & 17.1 & 1.14 \\ \hline
\end{tabular}%
\end{center}
\par
\vskip 0.5cm
\caption{Fraction (in $\%$) of the flux-averaged cross section associated to
states of a given multipolarity with respect to the total flux-averaged
cross section, i.e. $\langle \protect\sigma \rangle_{J^{\protect\pi%
}}/\langle \protect\sigma \rangle_{tot}$. Results are given for
all positive and negative states having total angular momentum $J$
between 0 and 5. The first column gives the considered nucleus,
the second tells if the results correspond to low energy
beta-beams or to a conventional source ($DAR$ for the
decay-at-rest of muons). In the former case the neutrino fluxes
are those of $^{18}$Ne ions having Lorentz ion boost
$\protect\gamma$ between 6 and 14.} \label{tab:fluxavernu}
\end{table}

\begin{table}[tbp]
\begin{center}
\begin{tabular}{|c|c|cccc|cccc|}\hline
$~^{A}$X~ &  &$~0^{+}~$ & $~1^{+}~$ & $~2^{+}~$ & $~3^{+}~$ &
$~0^{-}~$ &$~1^{-}~$ & $~2^{-}~$ & $~3^{-}~$\\\hline
$^{56}$Fe & 6 & 0.70 & 86.1 & 0.37 &0.33 & 0.37 & 6.83 & 5.32 & 0.0 \\
& 10 & 0.74 & 51.3 & 3.18 & 2.49 & 0.96 & 26.3 & 14.5 & 0.28 \\
& 14 & 1.01 & 31.0 & 8.95 & 5.61 & 0.68 & 33.9 & 15.7 & 1.61 \\
$^{208}$Pb & 6 & 0.57 & 7.04 & 11.8 & 3.82 & 7.21 & 49.9 & 18.7 & 0.60 \\
& 10 & 1.33 & 12.5 & 24.9 & 10.6 & 2.25 & 30.4 & 9.80 & 4.24 \\
& 14 & 1.91 & 13.6 & 25.2 & 9.48 & 0.56 & 18.3 & 9.19 & 10.6 \\ \hline
\end{tabular}%
\end{center}
\par
\vskip 0.5cm
\caption{Same as Table~\protect\ref{tab:fluxavernu},
but for the anti-neutrino nucleus cross sections and the
anti-neutrino fluxes produced by boosted $^{6}$He ions. }
\label{table:fluxaveranu}
\end{table}

\section{Conclusions}

We have presented new results on charged-current neutrino-nucleus
interactions. The theoretical approach used is the
random-phase-approximation among quasi-particles. Four nuclei of
particular interest have been chosen as typical examples, namely
oxygen, iron, molybdenum, and lead. We have given both the total
cross sections as a function of neutrino energy and the
flux-averaged cross sections associated to possible future
facilities producing low energy neutrino beams. These are either
based on conventional sources or on low energy beta-beams. A
detailed analysis of the states of different multipolarity
(allowed and forbidden) contributing to the cross sections has
been performed, as a function of the Lorentz ion boosts. The main
goal has been to explore if by combining different neutrino ion
accelerations, information on the various multipoles can be
extracted. We conclude that by varying the $\gamma$ factor, the
role of the spin-dipole states becomes as important as the
Isobaric Analogue and Gamow-Teller states allowing nuclear
structure studies of the forbidden states.

\vspace{.4cm}

We acknowledge the financial support of the EC under the FP6
"Research Infrastructure Action - Structuring the European
Research Area' EURISOL DS Project, Contract Number 515768 RIDS.

\end{document}